# A New Procedure for Microarray Experiments to Account for Experimental Noise and the Uncertainty of Probe Response

Alex E Pozhitkov*[a,b], Peter A Noble[b,c], Jarosław Bryk[a,d] and Diethard Tautz[a]

**Author affiliation:**

[a]Max-Planck-Institut für Evolutionsbiologie, August-Thienemann-Str 2, 24306 Plön, Germany

[b]Department of Periodontics, University of Washington, Seattle, WA, USA 98195.

[c]Ph.D Microbiology Program, Department of Biological Sciences, Alabama State University, 1627 Hall St. Montgomery, AL 36104

[d]Current address: National Centre for Biotechnology Education, University of Reading, Reading RG6 6AU, United Kingdom

***Corresponding author**

*AEP: alexander.pozhitkov@evolbio.mpg.de

Phone: +49-4522-763-283

short title: A New Microarray Experimental Design

## ABSTRACT

Although microarrays are routine analysis tools in biomedical research, they still yield noisy output that often requires experimental confirmation. Many studies have aimed at optimizing probe design and statistical analysis to tackle this problem. However, less emphasis has been placed on controlling the noise inherent to the experimental approach. To address this problem, we investigate here a procedure that controls for such experimental variance and combine it with an assessment of probe performance. Two custom arrays were used to evaluate the procedure: one based on 25mer probes from an Affymetrix design and the other based on 60mer probes from an Agilent design. To assess experimental variance, all probes were replicated ten times. To assess probe performance, the probes were calibrated using a dilution series of target molecules and the signal response was fitted to an absorption model. We found that significant variance of the signal could be controlled by averaging across probes and removing probes that are nonresponsive. Thus, a more reliable signal could be obtained using our procedure than conventional approaches. We suggest that once an array is properly calibrated, absolute quantification of signals becomes straight forward, alleviating the need for normalization and reference hybridizations.

**Keywords:** microarray calibration, noise, Freundlich isotherm, Langmuir isotherm, probe response, copy number variation

## PRACTICAL SUMMARY

Microarray signal intensities are not directly proportional to the concentration of their respective targets. Finding a relationship between signal intensity, target concentration and probe-target duplex sequence has been an intractable problem (10, and refs therein). Multiple convoluted statistical approaches have been proposed to alleviate the problem, but none of them have been established on simple approaches common in analytical chemistry such as calibration of sensors. We propose the following methodology:

1. Prepare a dilution series of a mRNA or gDNA biological sample.
2. Hybridize 6-8 microarrays with the series.
3. Each probe response fit to a curve, e.g., Langmuir, Freundlich, etc.
4. Remove probes exhibiting poor fits.
5. Hybridize sample of interest to a microarray
6. Calculate concentrations of the target based on the calibration curves.

## INTRODUCTION

Microarrays have been extensively used for examining gene expression and for detecting single nucleotide polymorphisms (SNPs) or copy number variations (CNVs) in genomic DNA (1,2). Yet, despite the general use of this technology, uncertainty remains in the interpretation of the array output. For example, several studies have shown that about 20 to 30% of expressed genes are identified as either up- or down- regulated solely depending on the algorithm used (3-5). Furthermore, it is currently routinely required to apply additional experimental tests, such as quantitative PCR, to verify results obtained from microarrays.

We argue that one reason for the uncertainty in the interpretation of the array output is insufficient measurement of experimental noise in current protocols. In the earlier array platforms (spotted arrays), the noise problem was mostly due to uneven surfaces of arrays and variability between arrays (e.g., ref. 6). This problem is now largely solved, partly because manufacturing of arrays has significantly improved and because internal quality checks are routinely implemented to account for this problem. On the other hand, any quantitative measurement is associated with measurement errors and even for a perfectly manufactured array, a determination of this error is expected to raise the statistical confidence in the measurement. However, an assessment of this error has so far not been implemented in the experimental procedures of microarray hybridization.

Another problem for the optimal design of arrays is the uncertainty of probe binding behavior. Although many parameters have been identified that affect probe binding behavior (7,8,9,10,11,12), it remains a challenge to design arrays in a way that makes probe binding behavior predictable. In high density arrays, such as Affymetrix, this problem is partly solved by summarizing results across multiple probes for the same target [e.g., Affymetrix probe sets, www.affymetrix.com]. Although this yields a major improvement in signal reliability, it is nonetheless still an inherently noisy procedure, since poorly responding probes may influence the signal in unpredictable ways. The alternative is validation and calibration of probes and we explore this here.

Our revised design for microarray experiments includes an estimate and control of experimental noise, as well as calibration of probes with a biological sample. Specifically, the calibration of probes allows one to identify poorly responding probes and subsequently remove them from the analysis. In addition, calibration allows one to directly determine target concentrations in biological samples from signal intensity, without the need to use reference hybridizations. To show that these procedures can definitely improve the accuracy of quantitative measurements using microarrays, we use two types of test arrays: one with short (25mer) probes and another with long (60mer) probes. Using test hybridization and adjusted statistical procedures, we show that a major improvement of signal reliability can indeed be obtained.

# MATERIALS AND METHODS

## Array experiments

Two custom arrays were used to test the new procedure. The first array (henceforth designated "25mer array") consisted of 5,912 (each replicated ten times) 25mer probes representing genes from the mouse X-chromosome that were taken from the Affymetrix "GeneChip Mouse Genome 430 2.0 Array". The second array (henceforth designated as "60mer array") consisted of 4,614 (each replicated 10 times) 60mer probes designed by Agilent to trace regions of structural variation, such as copy-number variation (CNV) in the mouse genome. All probes in both arrays were placed in random locations to allow the determination measurement error. The 25mer array was used to test the general utility of the approach for DNA and RNA hybridizations. The 60mer array was used to compare the performance of the Agilent standard procedure for CNV discovery to our protocol.

Genomic DNA (gDNA) and RNA was labeled according to the manufacturer's recommended protocol (Agilent). For the gDNA and RNA dilution series experiments (Figure 1), several samples of the recommended concentration were independently labeled. For gDNA, the labeled products were pooled together, precipitated with sodium acetate, and the resulting pellet was dissolved in Tris-EDTA buffer (10 mM Tris, 1 mM EDTA, pH 8.0). Then, the concentration of the DNA was measured with NanoDrop (Agilent Technologies), and a dilution series was prepared. For RNA, the yield was sufficient to make a dilution series by mixing several 5 µl aliquots of the independently labeled products and diluting the mix accordingly. Hybridization was conducted in the Agilent hybridization buffer at 48°C for approx. 17 h followed by recommended washing.

**Probe calibration** We visually inspected several hundred isotherms from the 25mer and 60mer arrays. Depending on the probe, it was possible to model the isotherms using Freundlich (13) or Langmuir (14-22) equations. We have then devised an automated algorithm to pick the most appropriate model[1] and applied it accordingly. The Langmuir equations is:

$$y = \frac{y_{\max} K x}{1 + K x} \qquad \text{Equation 1}$$

where $y$ is signal intensity; $x$ – concentration; $K$ – binding constant; $y_{max}$ – saturation level.

The Freundlich absorption equation is:

[1] http://web.evolbio.mpg.de/~alexander.pozhitkov/microarray123/

$y = ax^b$ Equation 2

where *y* is signal intensity; *x* – concentration; *a, b* – empirical parameters.

## Error calculation

The purpose of determining the relative error of the mean signal intensity *Err(y)* is to determine the relative error associated with calculated concentration, *Err(x)*. If the *Err(x)* is less than the acceptable level (e.g., 20%) than the calculated target concentration can be trusted. The *Err(y)* is calculated according to the sampling distribution (23) as standard deviation of *y* divided by the square root of the number of replicates and divided by the average of *y*. *Err(x)* is dependent on the model (see Equation 5 and Equation 6).

**Error calculation for the Freundlich equation**. An assessment of the error for the calculated concentration from the calibration curves can be achieved as follows. Given the form of the calibration curve (Equation 1),

$$y = ax^b \Rightarrow x = \left(\frac{y}{a}\right)^{\frac{1}{b}} \qquad \text{Equation 3}$$

one determines the relative error of *x* upon the error of *y*, by finding the differentials according to the error propagation theory (24). Assuming small uncertainties of the parameters *a* and *b*, which can be ensured by selecting calibration curves with a high goodness of fit (see below), the differentials are given by Equation 4.

$$dx = \frac{1}{b}\left(\frac{y}{a}\right)^{\frac{1}{b}-1}\frac{1}{a}dy \qquad \text{Equation 4}$$

The differentials are equivalent to standard deviations (24). Dividing both sides of the Equation 4 by the Equation 2 yields relative errors of *x* and *y*:

$$\frac{dx}{x} = \frac{1}{b}\frac{dy}{y} \Rightarrow Err(x) = \frac{1}{b}Err(y) \qquad \text{Equation 5}$$

**Error calculation for the Langmuir equation.** From the Langmuir equation, *x* (i.e., concentration) is obtained as follows:

$$x = \frac{y}{(y_{max} - y)K}$$

The error for calculated concentration is found according to the error propagation theorem (24).

$$dx = \frac{dy(y_{max} - y)K + yKdy}{(y_{max} - y)^2 K^2} = \frac{y_{max}Kdy}{(y_{max} - y)^2 K^2},$$

Where *dx* and *dy* are standard deviations of *x* and *y* respectively.

The relative error is as follows:

$$\frac{dx}{x} \equiv Err(x) = \frac{y_{\max} Kdy}{(y_{\max} - y)^2 K^2} \frac{K(y_{\max} - y)}{y} = \frac{dy}{y} \frac{y_{\max}}{y_{\max} - y};$$

Hence, Equation 6

$$Err(x) = \frac{y_{\max}}{y_{\max} - y} Err(y)$$

## Data analysis

Initially the data were stored and analyzed in an MS SQL database. We wrote three C++ programs to analyze the data for users. The programs can be downloaded at: http://web.evolbio.mpg.de/~alexander.pozhitkov/microarray123/. The probes and microarray data can be downloaded at: http://web.evolbio.mpg.de/~alexander.pozhitkov/data123/.

# RESULTS

## Measurement error

We assessed the extent of measurement error inherent in the standard microarray procedure by comparing the signals from ten replicated probes within each array. The arrays were hybridized with genomic DNA (gDNA) using the dilution series depicted in Figure 1. We observed indeed a general variance in the signal intensity among the 10 identical replicates of each probe. As an example, Figure 2A shows a typical case of signal intensities for a single 25mer probe at different dilutions. In this case, we observed up to four-fold differences for identical replicates. Averaging of the signal intensities of the 10 replicates, however, yields a good fit to a power-law function (Figure 2A). Hence, the variability in signal intensity of individual probes appears to reflect the measurement error.

The majority of probes have a variation coefficient of ~12 to 35%

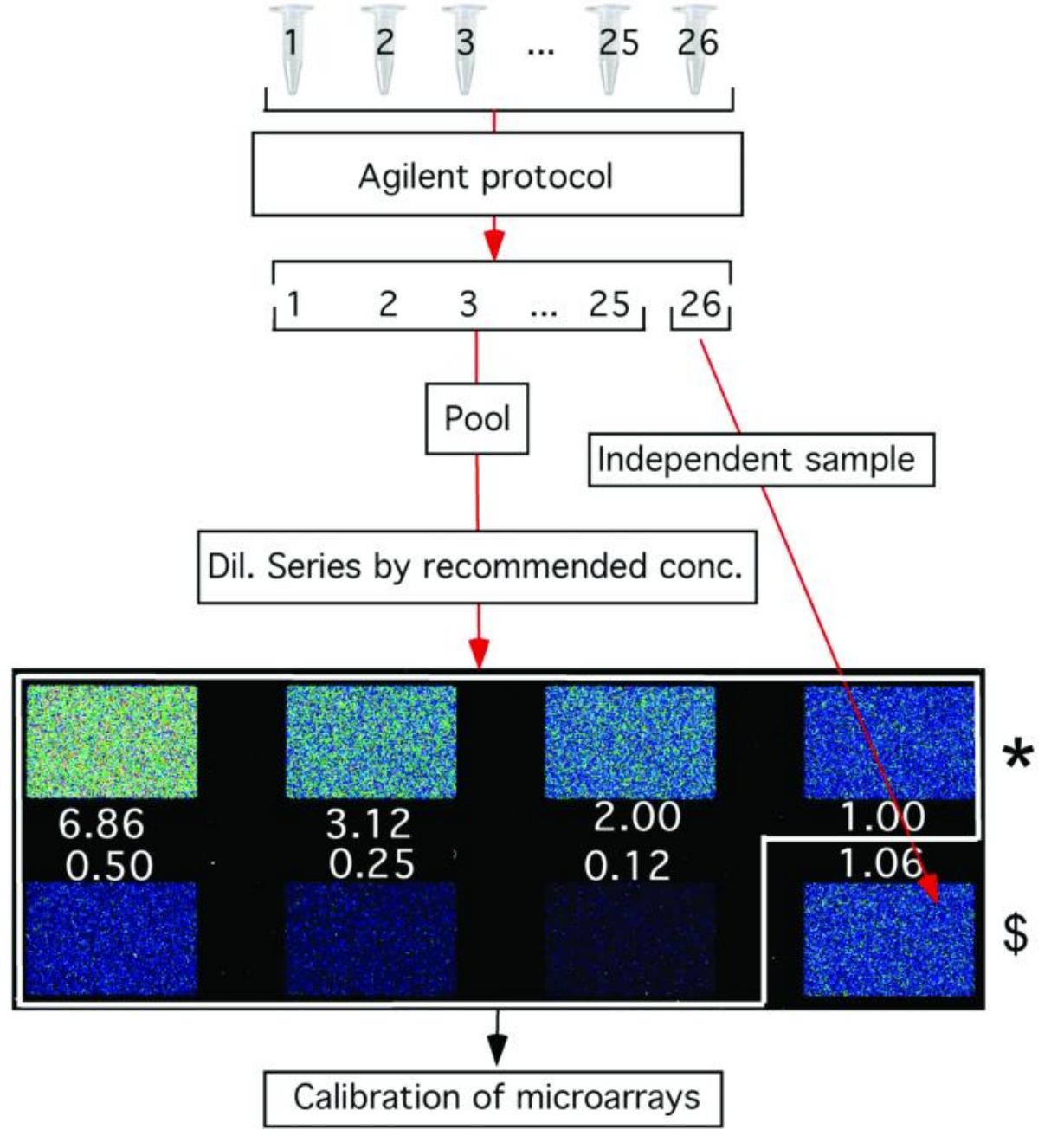

**Figure 1. Outline of the experimental design for recording a dose-response curve for each probe on an 8-plex Agilent microarray. The dilution series was created by pooling the labeled samples, serial diluting the pool, and hybridizing each diluted sample to the independent microarray. The arrays within the white box were used to calibrate the probes. One array marked with an asterisk (*) was used as a 'reference'. The independent 'test' array ($) is also shown. Numerical values indicate the target concentration in folds of the recommended concentration.**

for the 25mer array (Figure 2B) and ~7 to 10% for the 60mer array (Figure 2C), as assessed from the standard deviations of signal intensities over 10 replicates. Based on Figure 2B and C, the average CVs do not appear to depend on target concentrations. For a few probes on each array we observed unusually high CVs (data not shown). Inspection of these probes showed that this was always caused by technical problems (e.g., dust particles) in a single replicate.

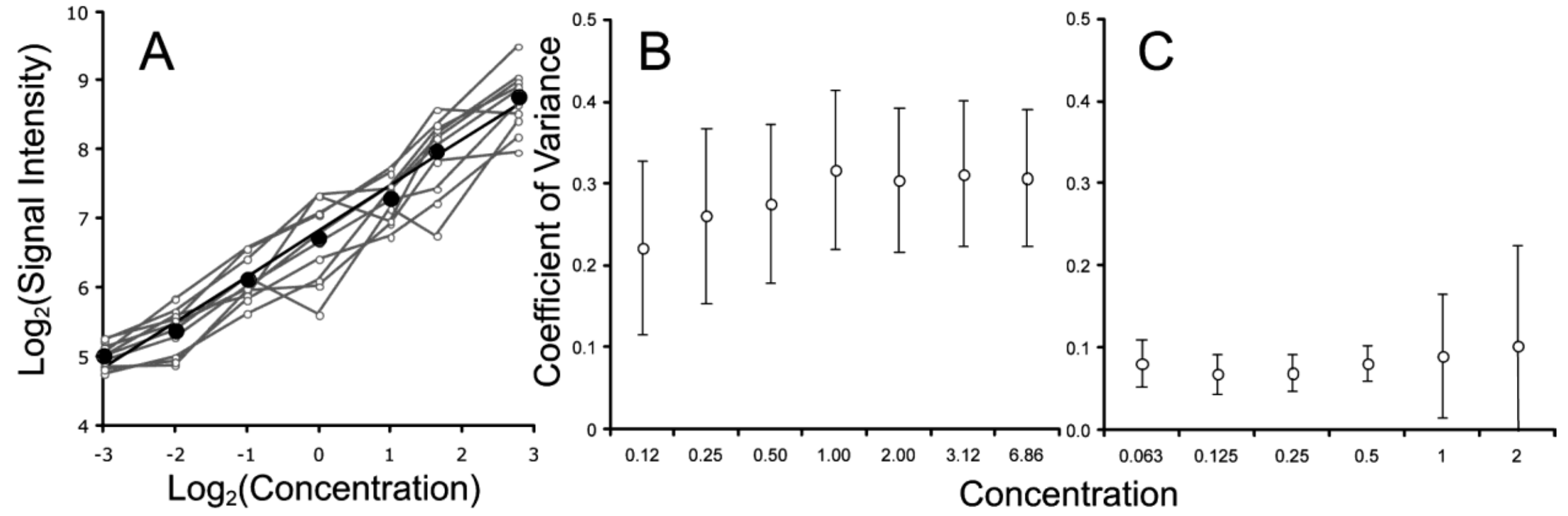

**Figure 2. Signal variances between replicated probes. (A) Typical Agilent array isotherms obtained using a dilution series of genomic mouse DNA, BL6 strain for a single probe and its replicates. Raw data (gray) and predicted isotherm based on the average signal intensity (black). (B) Mean and standard deviation of CV across all probes at each concentration for 25mer array. (C) Same for the 60mer array.**

**Calibration of probe behavior** The calibration should be used to determine the probe response function (i.e., calibration curve) and to remove poorly responding probes. In order to obtain calibration parameters of each probe, one has to determine the respective equation parameters, e.g., $R^2$, $k$ and $Y_{max}$ for the Languir equation; $a$ and $b$ for the Freundlich equation. The parameter estimation is done by a linear regression of the linearized data, i.e, $x/y$ vs $x$ for Langmuir model and log($y$) vs log($x$) for Freundlich model. Probes with low $R^2$ values for either equation are unlikely to be reliable for actual measurements. We suggest that probes below a cutoff of $R^2 \leq$ 0.98 should be removed from further analysis, but this cutoff could be individually adjusted for each experiment. For our experiments, we found that 1092 probes (18%) fell below this cutoff for the 25mer array and 1124 probes (24%) for the 60mer array. For the remaining probes, we found that the majority (98%) showed a better fit with the Freundlich equation for the 25mer array, while for the 60mer array, 30% showed a better fit for the Freudlich equation and 70% for the Langmuir equation. We determined the parameter distributions and $R^2$ values for each probe for both equations on both arrays (Figure 3). The Langmuir parameters for the 25mer arrays are not shown in Figure 3 due to the small number of probes (<2%) that followed the Langmuir model.

In contrast to gDNA arrays (such as CNV arrays), expression arrays are usually hybridized with mRNA targets. The optimal labelling procedure for RNA involves a RNA polymerization step (25). Because the physicochemistry of DNA:DNA hybridization differs from that of DNA:RNA (26), we expected that a calibration with a RNA target to yield different results from the calibration with the gDNA. We tested this using the 25mer array since these probes were derived from an expression array. Averaging and calibration was done as described above for the gDNA target. Comparison of the $a$ and $b$ parameters of the Freundlich equation for each probe revealed little correlation between DNA and RNA (data not shown). Hence, separate calibrations are needed for RNA and DNA targets.

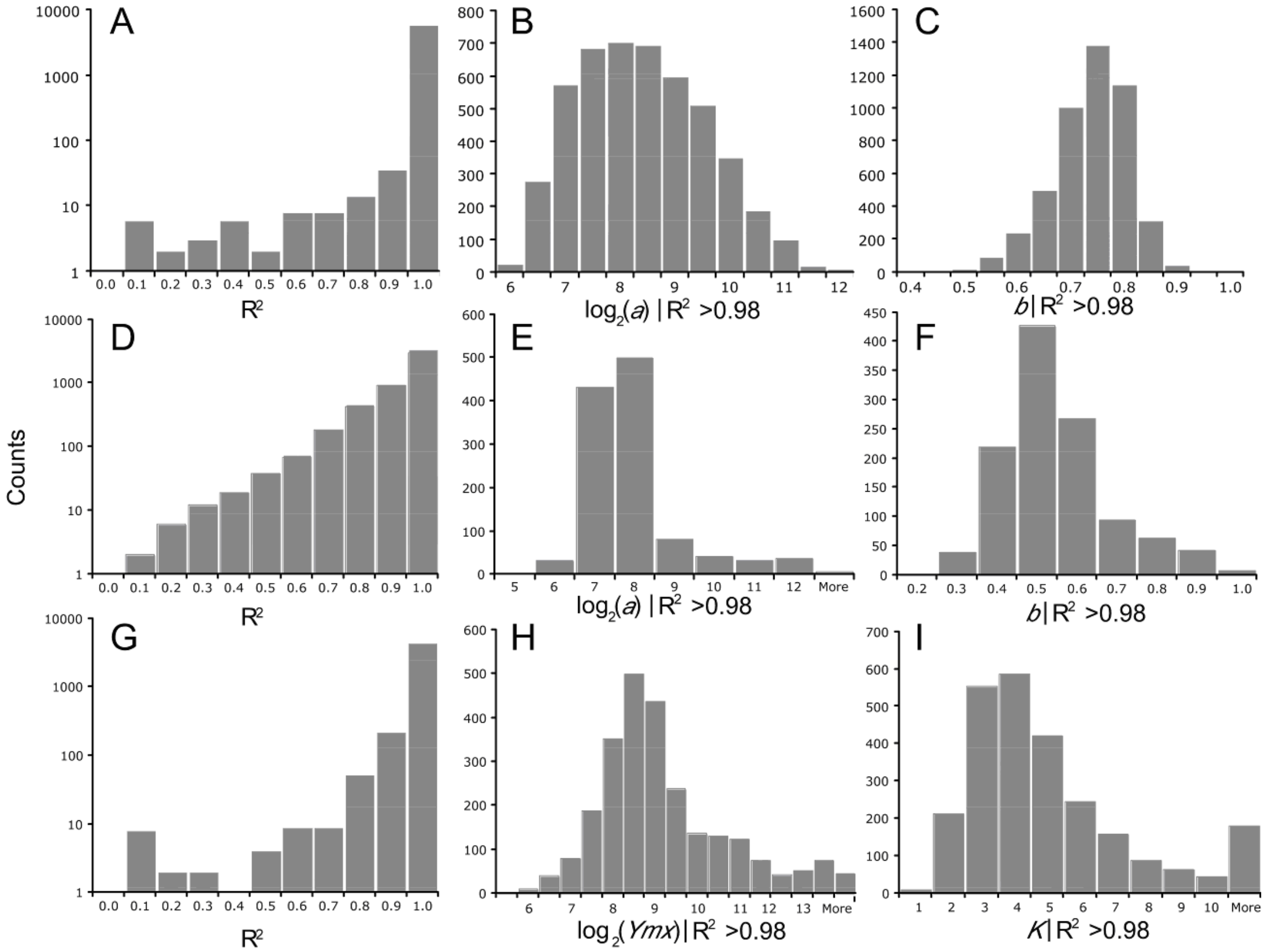

**Figure 3. Distribution of curve fitting parameters for the isotherm models. Panels A to F, Freundlich model, Panels G to I, Langmuir model. Panels A to C, 25mer arrays, Panels D to I, 60mer arrays. Panels A, D, and G: Distribution of $R^2$ across all probes. Panels B and E: Distribution of pre-exponent *a* for preselected probes. Panel C and F: Distribution of exponent *b* for preselected probes. Panel H: Distribution of Ymx for pre-selected probes. Panel I: Distribution of K for preselected probes.**

There is an additional problem with mRNA calibration because different mRNAs occur at different concentrations in a given sample. Specifically, probe signal intensities of mRNAs expressed at low levels (i.e., at low concentrations) will fall below the background level (Figure S1). Because of this problem, mRNA calibration should always be done in parallel to a given experiment in order to ensure appropriate representation of the mRNAs. Moreover, in contrast to the absolute calibration that is achieved for gDNA experiments, one can only determine a relative change in concentration for mRNA experiments since the concentration of the mRNAs in the calibration mix are not known.

**Assessing improvement of signal quality** In our first test we used identical DNA samples against each other (a reference and a test array, marked * and $ in Figure 1) in a gDNA hybridization experiment. The samples hybridized to these arrays were derived from the same DNA; therefore, no signal variation was expected and their signal intensity ratio should be equal to 0 in the $\log_2$ scale. Any deviation from 0 represents the noise in the experiment. Figure 4A and 4B show the ratios of signal intensities of the reference and test samples for individual probes versus ten averaged replicates, respectively. For the individual probes, only 62.3% of the

ratios spanned an acceptable range from 0 to 1.4 ($log_2$ values between -0.5 to +0.5; orange columns). For signal ratios of averaged probes, 94.2% were within this range. A similar improvement (88.9%) was obtained when we used calibration from averaged probes instead of ratios (Figure 4C). Although this is lower than the one obtained using the ratio of the averages, the calibration method is superior because it removed probes that have non-linear behavior. The calibration approach thus results in a highly symmetric distribution and indicates that more quantitative signal estimates are obtained.

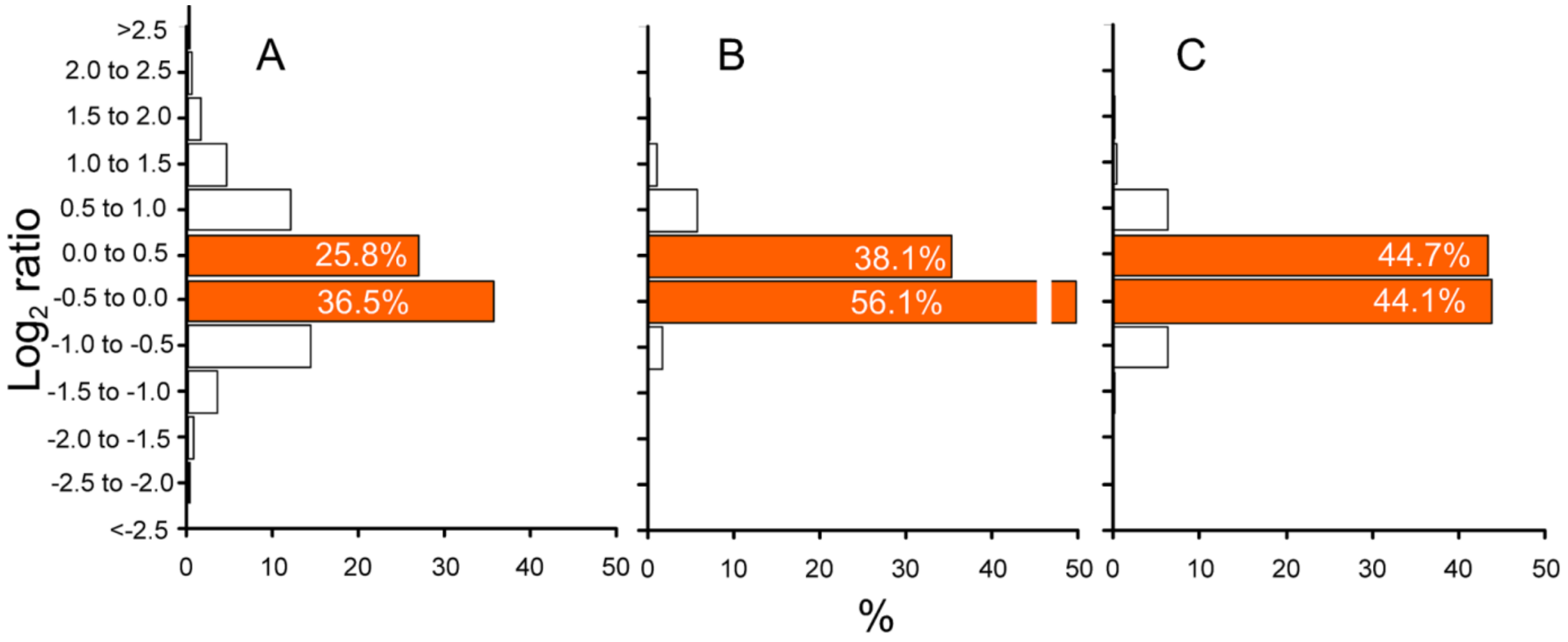

**Figure 4. Overall assessment of noise reduction using the same DNA sample. Since the sample is compared against itself, the $log_2$ ratio should be 0. Hence, all values above or below 0 are experimental noise. (A) No noise reduction - ratio of signal intensities between all individual probes (*n* = 5,912); (B) ratio of signal intensities from 10 averaged probes (*n* = 5,912); (C) concentration values from calibrated isotherms of all responsive probes ($R^2$ > 0.98), a value is included only if its relative error is under 20% (*n* = 4,406).**

The second test was aimed at assessing signal improvement in an actual experiment. Specifically, we compared the conventional analysis procedure using Agilent software to our calibration approach using a given CNV region in the mouse genome. The CNV analysed consisted of an approximate 5 kb fragment present in variable copy numbers between wild type individuals, but only one copy in the reference strain (C57B1/6). Figure 5 shows that the Agilent ratio analysis (upper panel) is indeed much noiser since many of the probes show values that are two to three standard deviations from the average mean ratio (red and blue dots). In contrast, the calibrated probes (lower panel) showed mostly a smooth distribution. Both methods detected the CNV in question (indicated by the blue bar at the bottom), but the copy number estimate is more reliable for the calibrated probes. This comparison suggests that our protocol can be expected to result in fewer false positive calls and a better measuring capacity in CNV studies.

**Number of replicated probes** Although the above experiments used 10 replicated probes for averaging, it would be of interest to know whether or not this is an optimal number. To address this question, we randomly selected 2 to 10 replicated probes from both 25mer and 60mer arrays and back-calculated the expected concentration of targets for the standard experiments (target concentration of 1x). The calculation was based on the calibration equations and parameters derived from 10 replicated probes because they are closest to the truth. As to be expeccted, we find a higher variance for estimating the true concentration when fewer probes were used (Figure 6). For the 25mer array, 10 probes produced the lowest variance, but the shape of the curve suggests that even more probes might be beneficial. For the 60mer array, we see no improvement beyond 6 replicates, i.e., this might be the optimal number of probes for this array type (Figure 6).

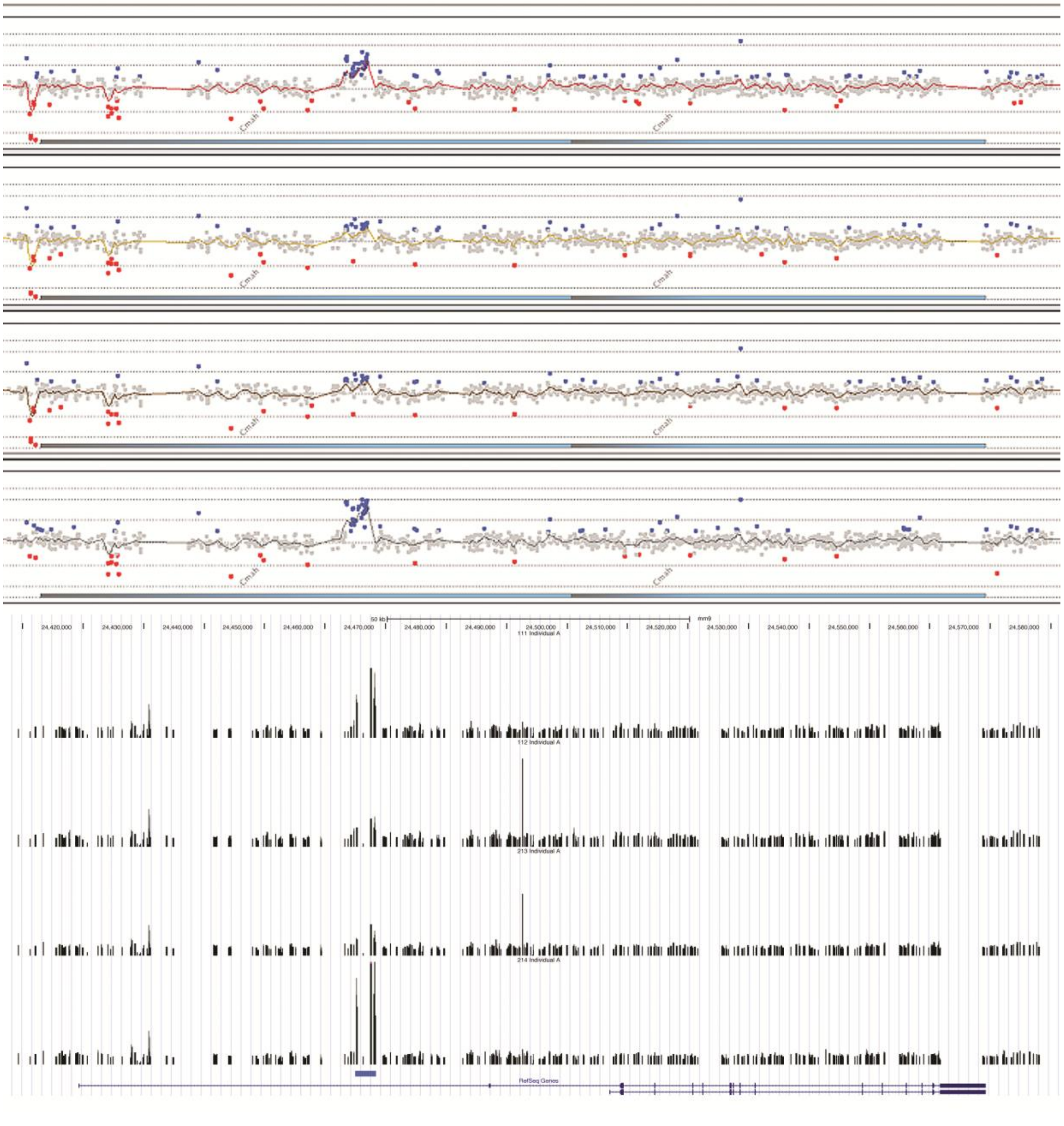

**Figure 5. Comparison of analysis procedures for copy number variation of a gene region in mice. Four different wild type mice were analyzed, each represented as a track. Top: output from the ratio analysis implemented in the Agilent software (ratio with respect to DNA from an C57Bl/6 inbred mouse strain). The input was the concentrations derived from the ten averaged probes, but without calibration and removal of non-responding probes. Colored dots represent values larger (blue) or smaller (red) then $\log_2 = 0.5$. Bottom: concentration calculations based on the method proposed here, non-responding probes removed (> 20% error in any of the experiments on the array). The values were normalized with respect to average intensities on the array and displayed as custom track in the UCSC genome browser.**

**Variance in the sample preparation** Although the averaging and calibration removed much of the noise, a known addition source of noise comes from target preparation. Specifically, the target fragmentation and labelling procedures involve several enzymatic steps (i.e., PCR enzymatic digestion), which have previously been reported to introduce variability (27). Figure 7 shows that the noise is indeed clearly higher in the test sample (16.1 % of comparisons outside the acceptable range) than in the pooled sample (5% outside). This result supports the notion that preparing multiple independent samples and then pooling the preparations can reduce the variance inherent in the target sample preparation.

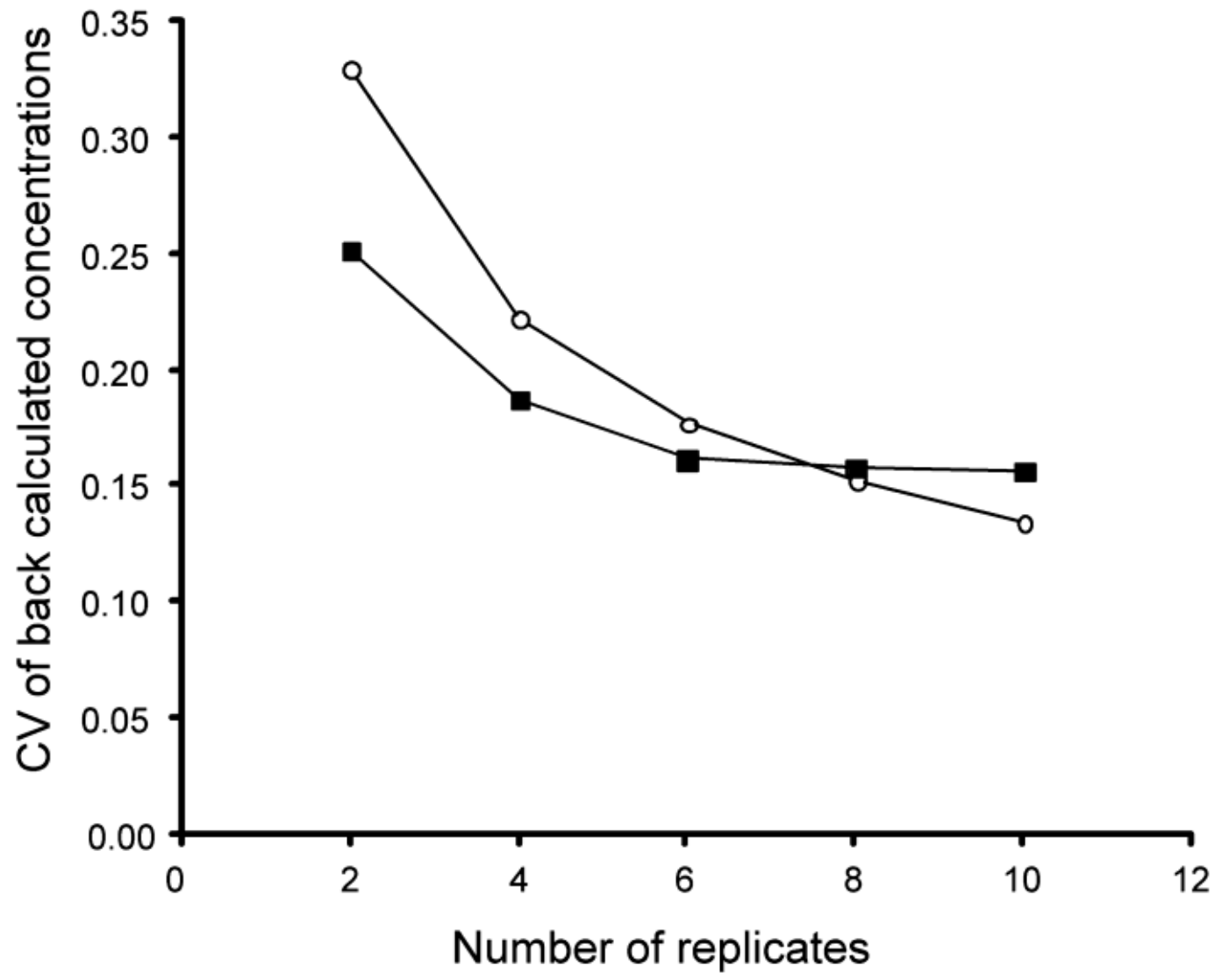

**Figure 6. Coefficient of variation (CV) decrease for estimating the true concentration and dependence of probe replication. Circles, 20mer array; Squares, 60mer array.**

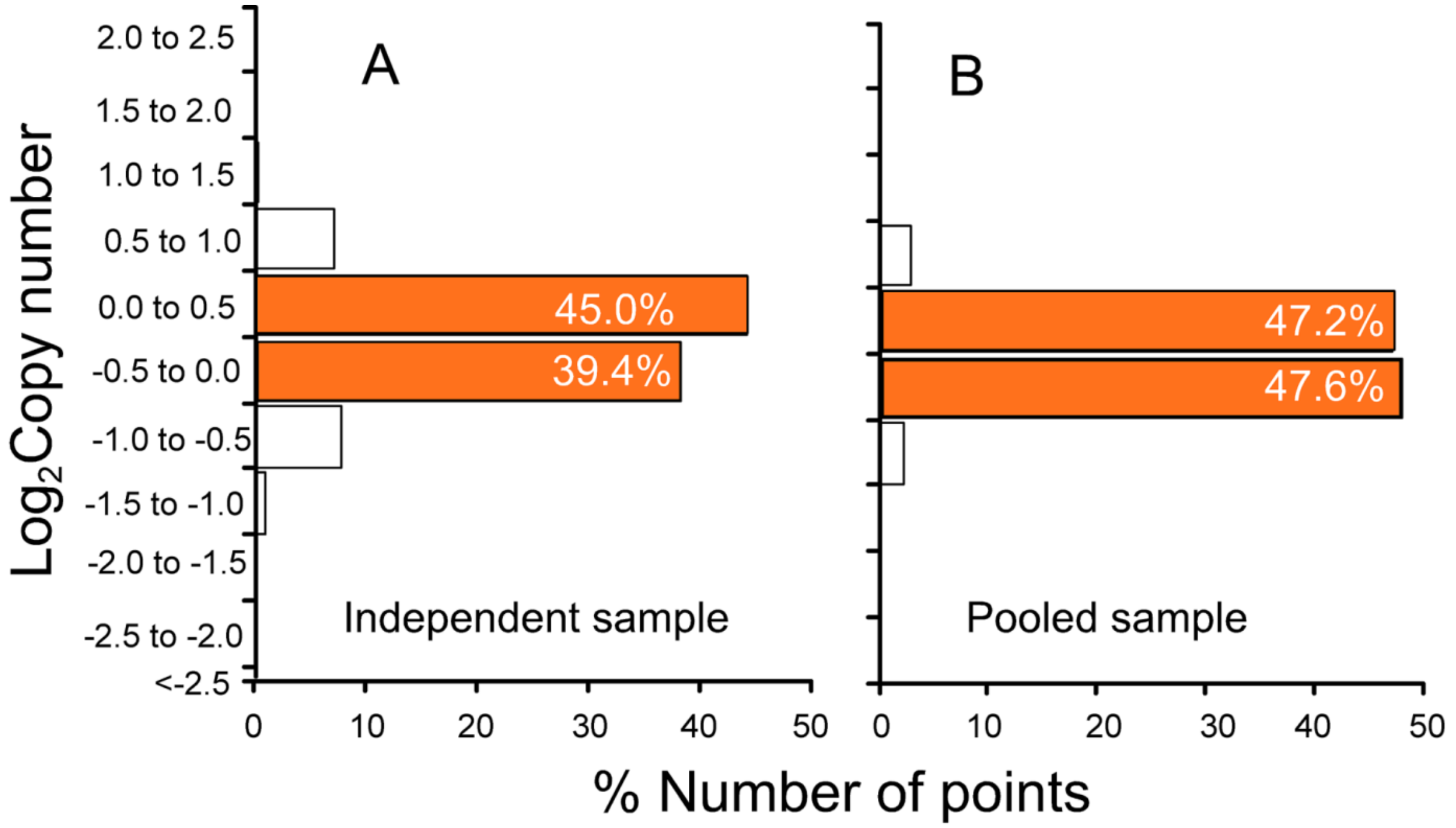

**Figure 7. Comparison of probe labeling effects on noise reduction. (A) Sample $ (*n* = 4,775) and (B) sample * (*n* = 4,767) from Figure 1. Note that the calibration was conducted without sample *. Ratios were calculated from calibration curves ($R^2$ > 0.98) and a value is included into the histogram only if its relative error is under 20%.**

## DISCUSSION

Our procedure is based on a common approach used in physics and analytical chemistry to experimentally determine the performance of a sensor (i.e., probe) and the magnitude of a measurement error. Once the measurement error is known, simple statistics can be used to obtain estimates of the true values. Knowing the error distribution, one can also reject outliers. We have conjectured and experimentally verified that there is indeed such an error distribution at the level of probe binding and target preparation (the labeling procedure).

We showed that the fidelity of estimating the true target concentration increased significantly with multiple replications of the probes and that this fidelity was dependent on probe length. For 25mer arrays, at least 10 probes seem to be necessary, while 6 probes seem to be sufficient for the 60mer array. Replicating probes limits the number of different probes that can be surveyed on an array. This presents a trade-off between the quality of signal and the number of different genes that can be studied, at least for array designs that can not compensate this with very high probe densities.

For controlling the error associated with target preparation, it is not strictly necessary to run a separate experiment for each independent labeling since the exact error is not relevant for subsequent calculations. Instead, one can perform independent labeling reactions with the same target mixture. This results in an averaging of the error in the same way as one would do statistically, although the magnitude of the error remains unknown.

Using the biological sample itself for calibration entails the risk that one is not only calibrating for the specific signal, but also for any unwanted nonspecific hybridization. The problem of cross-hybridization by similar target sequences can usually be addressed by applying algorithms in the probe design phase, provided full genome information is available. It remains a problem, though, that the total signal intensity contains specific and nonspecific hybridization signal and this will be probe-specific. Hence a remedy would be to design more than one probe for a given region (e.g., a gene) and compare the signals. Those probes that report the same dynamics are most likely unaffected by nonspecific hybridization. Ideally, one should calibrate an array with pure target sequences to precisely determine the extent of nonspecific hybridization and remove those probes that are affected by it. However, this will be expensive and may eventually only be feasible for diagnostic arrays that are used in large numbers.

## CONCLUSION

Due to the uncertainties of probe binding behavior, conventional array procedures use ratio of signal intensities between “control” and “treatment” samples. However, the noise associated with the individual probe signal intensity in the control and in the treatment could potentially make the problem even greater, because the noise is additive. With our procedure, one gets an absolute quantification of targets, without reference to a control. In addition, the calculation of the target concentration from the calibration curve takes care of the non-linearity of the signal. Thus, a two-fold change in calculated concentrations represents *a true two-fold change*. This is not the case for the conventional array procedures. For instance, a two-fold change means only that signal intensities differ by two-fold with an unknown difference in the true target concentrations. Therefore, calibrated microarrays offer a significant improvement in the quantification of targets. In addition, calibrated microarrays are more economical than conventional arrays because once they are calibrated no further reference hybridizations are necessary.

## ACKNOWLEDGEMENTS

We thank E. Blom-Sievers for technical support and the members of the laboratory for discussions. Discussions among the participants of the MPI international conference entitled "Physicochemical fundamentals of DNA hybridization on surfaces as applied to microarrays and bead-based sequencing technologies" at Ploen, Germany on May 9 to 12, 2011(http://web.evolbio.mpg.de/ploenworkshop/) also contributed to the content of the manuscript.

## FUNDING

This work was supported by institutional funds of the Max-Planck Society and by NSF grant DBI 1038671.

*Conflict of interest statement.* None declared.

## Authors Contributions

AP, PAN, DT developed the new design; AP, JB carried out the experiments; AP, PAN, JB carried out the analysis; AP, PAN, DT wrote the manuscript.

# REFERENCES

1. Fan,J.B., Chee,M.S., Gunderson,K.L. (2006) Highly parallel genomic assays. *Nat. Rev. Genet.,* **7,** 632-644.
2. Hoheisel,J.D. (2006) Microarray technology: beyond transcript profiling and genotype analysis. *Nat. Rev. Genet.,* **7,** 200-210.
3. Millenaar,F.F., Okyere,J., May,S.T., van Zanten, M., Voesenek,L.A., *et al.* (2006) How to decide? Different methods of calculating gene expression from short oligonucleotide array data will give different results. *BMC Bioinform.* **7**, 137.
4. BarashY., Dehan,E., Krupsky,M., Franklin,W, Geraci M., *et al.* (2004) Comparative analysis of algorithms for signal quantitation from oligonucleotide microarrays. *Bioinform.* **20**, 839-846.
5. Seo,J., Bakay,M., Chen Y.W., Hilmer S., Shneiderman B., et al. (2004) Interactively optimizing signal-to-noise ratios in expression profiling: project-specific algorithm selection and detection p-value weighting in Affymetrix microarrays. *Bioinform.* **20**, 2534-2544.
6. Steger D., Berry D., Haider S., Horn M., Wagner M., et al. (2011) Systematic spatial bias in dna microarray hybridization is caused by probe spot position-dependent variability in lateral diffusion. *PLoS ONE* **6**, e23727.
7. Held G.A., Grinstein G., Tu Y. (2006) Relationship between gene expression and observed intensities in DNA microarrays–a modelling study. *Nucl. Acids Res.* **34**, e70.
8. Matveeva O.V., Shabalina S.A., Nemtsov V.A., Tsodikov A.D., Gesteland R.F., et al. (2003) Thermodynamic calculations and statistical correlations for oligo-probes design. *Nucl. Acids Res.* **31**: 4211-4217.
9. Mueckstein U., Leparc G.G., Posekany A., Hofacker I., Kreil D.P. (2010) Hybridization thermodynamics of NimbleGen Microarrays. BMC Bioinformatics **11**, 35.
10. Pozhitkov A.E., Boube I., Brouwer M.H., Noble P.A. (2010) Beyond Affymetrix arrays: expanding the set of known hybridization isotherms and observing pre-wash signal intensities. *Nucl. Acids Res.* **8**, e2.
11. Pozhitkov A.E., Stedtfeld R.D., Hashsham S.A., Noble P.A. (2007) Revision of the nonequilibrium thermal dissociation and stringent washing approaches for identification of mixed nucleic acid targets by microarrays. *Nucl. Acids Res.* **35**, e70.
12. Pozhitkov,A., Rule,R.A., Stedtfeld,R.G., Hashsham,S.A. and Noble,P.A. (2008) Concentration-dependency of nonequilibrium thermal dissociation curves in complex target samples. *J Microbiol. Methods.* **74**, 82-8.
13. FreundlichH. (1932) On the adsorption of gases. Section II Kinetics and energetics of gas adsorption. *Trans Faraday Soc.* **28** 195-201.
14. Burden,C., Pittlekow,Y., and Wilson,S. (2006) Adsorption models of hybridisation and post-hybridisation behaviour on oligonucleotide microarrays. *J. Physics: Cond. Mat.* **18**, 5545-5565.
15. Li,S., Pozhitkov,A., Brouwer,M. (2008) A competitive hybridization model predicts probe signal intensity on high density DNA microarrays. *Nucl. Acids Res.* **36**, 6585-6591.
16. OnoN., Suzuki,S., Furusawa,C., Agata,T., Kashiwagi,A., *et al.* (2008) An improved physico-chemical model of hybridization on high-density oligonucleotide microarrays. *Bioinform* **24,** 1278-1285.
17. Burden C.J. (2008) Understanding the physics of oligonucleotide microarrays: the Affymetrix spike-in data reanalysed. *Phys. Biol.* **5,** 016004.
18. Halperin,A., Buhot,A., and Zhulina,E.B. (2004) Sensitivity, specificity, and the hybridization isotherms of DNA chips. *Biophys. J.* **86,** 718-730.
19. Halperin,A., Buhot,A., and Zhulina,E.B. (2006) On the hybridization isotherms of DNA microarrays: the Langmuir model and its extensions. *J. Phys. Condens Mat* **18**, S463–S490.
20. Heim,T., Tranchevent,L.C., Carlon,E., Barkema,G.T. (2006) Physical-chemistry-based analysis of Affymetrix microarray data. *J. Phys. Chem. B.* **110**, 22786-22795.
21. Hekstra,D., Taussig,A.R., Magnasco,M, and Naef,F. (2003) Absolute mRNA concentrations from sequence-specific calibration of oligonucleotide arrays. *Nucl. Acids Res.* **31**,1962-1968.
22. Held,G.A, Grinstein,G., and Tu,Y. (2003) Modelling of DNA microarray data by using physical properties of hybridization. *Proc. Nat. Acad. Sci. U S A* **100**, 7575.
23. Miller,M., and Miller,J.E. (2004) Freund's Mathematical Statistics with Applications, 7th ed. Prentice Hall, NJ.
24. Meyer,S.L. (1975) Data Analysis for Scientists and Engineers, Wiley, NY.
25. StudierF.W., Rosenberg,A.H., Dunn,J.J., and Dubendorff,J.W. (1990) Use of T7 RNA polymerase to direct expression of cloned genes. *Methods Enzymol.* **185**, 60-89.
26. Lang B.E. and Schwarz F.P. (2007) Thermodynamic dependence of DNA/DNA and DNA/RNA hybridization reactions on temperature and ionic strength. *Biophys Chem.* **131**, 96-104.

27. Osborn,A.M., Moore,E.R., and Timmis,K.N. (2000) An evaluation of terminal-restriction fragment length polymorphism (T-RFLP) analysis for the study of microbial community structure and dynamics. *Environ Microbiol* **2,** 39-50.

# SUPPLEMENTARY INFORMATION

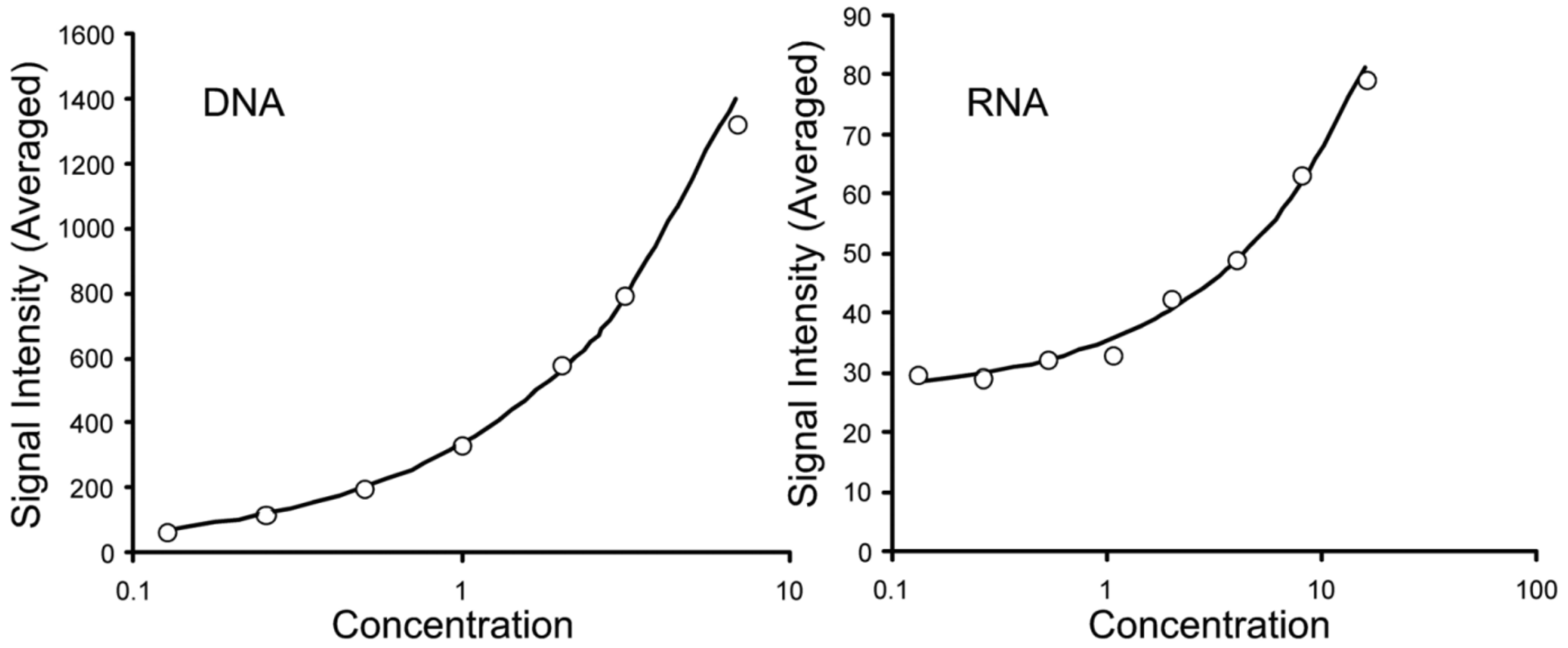

**Figure S1. Comparison of isotherms for a given probe (average values from 10 replicates). Left: hybridized to DNA ($R^2$=0.99); right: hybridized to RNA ($R^2$=0.99), the offset value c for this sample is ~26.**